\documentclass[manuscript, nonacm]{acmart}
\AtBeginDocument{%
  }

\definecolor{gray}{HTML}{808080}
\definecolor{teal}{HTML}{21908C}
\definecolor{yellow}{HTML}{FDE725}
\definecolor{blue}{HTML}{1E2BC5}
\definecolor{purple}{HTML}{5D3FD3}
\definecolor{green}{HTML}{238E12}

\newtoggle{comments}
\toggletrue{comments}

\title{What's so Human about Human-AI Collaboration, Anyway? Generative AI and Human-Computer Interaction}

\author{Elizabeth Anne Watkins}
\email{elizabeth.watkins@intel.com}
\author{Emanuel Moss}
\email{emanuel.moss@intel.com}
\author{Giuseppe Raffa}
\email{giuseppe.raffa@intel.com}
\author{Lama Nachman}
\email{lama.nachman@intel.com}
\affiliation{%
  \institution{Intel Labs}
  \city{Hillsboro}
  \state{Oregon}
  \country{USA}}

\begin{document}
\begin{abstract}

While human-AI collaboration has been a longstanding goal and topic of study for computational research, the emergence of increasingly naturalistic generative AI language models has greatly inflected the trajectory of such research. In this paper we identify how, given the language capabilities of generative AI, common features of human-human collaboration derived from the social sciences can be applied to the study of human-computer interaction. We provide insights drawn from interviews with industry personnel working on building human-AI collaboration systems, as well as our collaborations with end-users to build a multimodal AI assistant for task support.
\end{abstract}

\maketitle


\section{Introduction}

While still a relatively recent phenomenon in the history of artificial intelligence, generative AI has been proposed as a particularly powerful tool through which humans and computers can interact~\cite{bernstein2023architecting}. The startlingly rapid development of generative AI interfaces has been accompanied by sizeable investments in their deployment~\cite{rees_future_2025} to mediate a vast range of human interactions with not only computers, but also other human beings and social institutions~\cite{zittrain_we_2024}. Given how central generative AI approaches may become for such a wide range of interactions~\cite{fui-hoon_nah_generative_2023} and how "human-like" such interactions are posited to be ~\cite{obrenovic_generative_2024},\footnote{The extent to which generative AI interactions \textit{are} or \textit{are not} human-like is a separate question. For this paper, the \textit{claim} that generative AI is capable of producing human-like interactions serves as motivation.} \textbf{this paper explores the features of human-human interactions that are relevant for understanding, designing, analyzing, and improving human-AI collaboration}. The technological affordances~\cite{meredith_analysing_2017, gaver_technology_1991} of generative AI make these features of human-human collaboration particularly instructive for the design of human-computer interactions.

Below, we lay out five general features that characterize human-human interaction that are relevant for human-AI collaboration. These are important for developing generative AI-based approaches to interaction that are in line with human expectations for these types of interactions\footnote{In this work, we use the term "collaboration" to refer to the project, whereas the term "interaction" refers to the exchanges between human-human or human-AI.}: 1) Indeterminacy, 2) Contextual Integrity, 3) Contextual Controls, 4) the triad of Trust, Mistrust, \& Vulnerability, and 5) Translation. This relevance matters because, as we observe in our empirical analysis, human expectations of collaboration also dictate their expectations of human-AI collaboration. In this paper, we detail empirical examples where we saw that human expectations of human-AI collaboration were not met, leading to friction, misinterpretation, and frustration.

The five general features that characterize human-human interaction relevant for human-AI collaboration were identified through a series of interviews conducted with personnel working in an industry research lab on multiple overlapping projects focused on ``human-AI collaboration". These projects were premised on the idea that humans and AI systems, interacting with each other, may be able to work together to accomplish a task or achieve a goal that would have been more difficult or impossible without such collaboration. For example, an AI system might be able to learn a task well enough from interacting with human experts at that task to subsequently support those experts in ways that improve their performance at that task. How researchers understood ``collaboration" and ``interaction", therefore, were crucial starting points for identifying the features of human interaction that could be leveraged for human-AI collaboration.

Interviews were semi-structured. They following a loose script intended to elicit reflections from technologists on their explicit goals for human-AI collaboration research within the context of their specific project, and on how they imagine such collaborations might unfold through human-AI interactions. Most of these projects focused on performance or task support, where an AI system, whether in a computer interface or embodied in a robotic apparatus, provided assistance to a human completing a task. These tasks ranged from physical work inside factories, such as cleaning a machine component or constructing a built object, to digital tasks such as analyzing data or writing code. A team of social scientists trained on conversational analysis, human-computer interaction, and social theory analyzed these interviews. They looked in particular for ways in which technologists' descriptions of collaboration could be mapped onto features of human interaction, features that have been robustly established through empirical social science. These features of human interaction hold potential value for the design of human-AI systems to streamline interactions, improve trustworthiness, improve user experience, and ultimately deliver greater uplift and impact to human activities across a range of metrics. We supplement these interviews with empirical data collected over two years of work developing a multimodal AI assistant for the manufacturing space. From observations, interviews, and participatory design exercises with end-users of this AI assistant, we draw deeper insights into how features of human interaction can be successfully operationalized for human-AI systems.

\section{Background Assumptions about Human-AI Interaction}
Each engineer we interviewed described ``collaboration" through specific assumptions about how humans and AI systems interact. These assumptions were either explicitly articulated, or implicit in how engineers described their individual projects, and were derived from a grounded analysis of interview transcripts~\cite{pidgeon1991use, glaser2017discovery}. For brevity, they are briefly discussed here. The most common assumption was that human-AI interactions are directly analogous to how one human might interact with another. This analogy was seldom elaborated upon. Rather, human-human interaction was referenced in aspirational terms. Engineers, when listing the many goals of their projects, described how they aspired to produce interactions that were similar to human-human interactions. This goal was echoed in diagrammatic descriptions of human-AI collaboration that enrolled humans and AI systems, in turns, as `senders' or `receivers' of information. This model underlies broad swaths of information theory, as articulated by Claude Shannon~\cite{shannon1948mathematical}, and corresponds in some respects with structuralist approaches to linguistics~\cite{saussure_course_2011} but contrasts with constructivist or performative approaches to linguistics~\cite{austin_how_1962}. 

Another prominent assumption about human-AI collaboration arose in how engineers described a common goal - for human-AI interactions to be ``natural", i.e., akin to human-human interaction. They described how they decided to apply design changes whenever they determined that an interaction between human and AI might be ``unnatural", i.e. annoying or burdensome to a user. For prior HCI literature, ``natural" interactions encompass a broader set of communicative modalities (e.g. gesture or facial expression in addition to voice or text)~\cite{bernsen2001natural} or input modalities (e.g. tactile interfaces instead of graphic interfaces)~\cite{liu_natural_2016}. However, for these engineers ``natural" and ``unnatural" interactions could be better understood as seamless \cite{chalmers2003seamful} or not, in terms of turn-taking between human and AI system or the capacity for interactions to occur while humans continued their existing set of tasks otherwise uninterrupted by the AI system's prompts. This assumption that interactions ought to be ``natural" was embedded in other background assumptions as well. Specifically, engineers described ``task execution" as a paradigmatic opportunity for human-AI collaboration, in which humans could instruct, train, or transmit intent through the execution of a task and would not need to substantially alter their routines to complete an interaction. Notably, the reciprocal was not true--AI systems were never described as being able to interact with humans through their own task execution.

These background assumptions are incompatible with each other in specific ways. Drawing on social theory about how humans interact with one another can help us outline these incompatibilities. For one, the formalized assumption that humans occupy a binary, static interaction of ``sending" and ``receiving" information by turns is not compatible with the assumption that interaction design should be "naturalistic" in nature. Sender and receiver operating characteristics are useful for a mathematical theory of communication but humans ``do" much more with language~\cite{austin_how_1962} than send and receive information. So, these two assumptions complicate each other, as will be discussed at greater length below. For another, the assumption that task execution or demonstration is somehow more natural if it does not interrupt human activity is complicated by the observation that interruption and the ``repair" ~\cite{jackson_rethinking_2014, meck2024failing, raudaskoski1990repair} or restoration of continuity is a crucial feature of so-called ``natural" human interactions, but also requires significant scaffolding and support, which will also be discussed at length below.

\section{Five Properties of Human-Human Interaction}

\subsection{Indeterminacy.} Human interactions take place within---and depend upon---significant amounts of indeterminacy. Interactions between humans seldom have a predetermined outcome or endpoint. Rather, the outcome of the interaction is emergent from the interaction. Many aspects of the interaction remain vague or indeterminate throughout~\cite{bazzanella2011indeterminacy}. This vagueness enables parties in an interaction to negotiate the outcome, as a clear statement of positions would not leave room for compromise or accommodation. Additionally,  an interaction may serve multiple goals simultaneously. This can be illustrated by mundane ``dinner conversations" in which the conversation itself is a ``jointly accomplished activity", at the same time that talking helps determine and reinforce social roles, share information, and accumulate social capital~\cite{gardner2004conversation}. Crucially, the goal of human interactions is often indeterminate throughout their duration and is mutually arrived at through the interaction itself ~\cite{barad2007meeting}. While the goal can be anticipated prior to an interaction or reconstructed in retrospect~\cite{garfinkel2023studies, suchman1987plans}, the interaction itself is indeterminate while it is underway. Designers of human-AI interactions can use this insight about the indeterminacy of interactions as a way to expand how both humans and AI systems are enrolled beyond that of ``sender" or ``receiver" of information and to account for the multiple tasks that might be accomplished through an interaction, in addition to the tasks for which the interaction is primarily designed. An AI system that enrolls experts to train a model through demonstrations of a task, for example, learns how to do the task through the interaction with experts. But that is not all that is accomplished. The system's error is reduced through successive expert demonstrations, certainly, but also the expert's expertise is reinforced by how they enrolled. Such an interaction is also rife with indeterminacy, as the efficacy of the demonstration for training the system is uncertain until the interaction is complete. 

\subsection{Contextual Integrity.} Originally developed to think about privacy, contextual integrity ~\cite{nissenbaum_privacy_2010} makes a normative claim about human interactions: that interactions happen within specific contexts and that, within those contexts, they are subject to norms that shape how interactions should transpire. Context is important for interactions because it helps establish those norms, which shift from one context to another. A dinner party has different interactional norms as a board meeting, which has different interactional norms than shopping at a grocery store. These norms need to be in place for information to flow properly or for the interaction to be successful. 
Interactions between humans are governed by contextual norms, and human-AI interactions can be, as well. Context is also key for expectations around knowledge, and how knowledge can be operationalized for inference and insights: contextual boundaries determined what's "in scope" and what's "out", and certain conclusions, insights, and recommendations should be known to be "out of scope" as determined by contextual boundaries. While human-AI collaborations often present novel challenges for interactions designers and participants alike, they often produce interaction styles that are analogous to human interactions and human-AI interaction norms can be adopted from those analogues. 

We can see the importance of contextual integrity in empirical observations. In our work developing a multimodal AI assistant with a participatory design approach, our team relied on end-users to provide data in several modalities, as well as data labeling, through demonstrating manufacturing techniques via a process called `bootstrapping'. During a \emph{visual} bootstrapping session (which followed a series of \emph{narrative} bootstrapping sessions) to help develop a new computer-vision system, the end-user was asked to “show the camera a bearing.” By that time, the end-user had already been working with our team for several months, and had already demonstrated his task to the cameras many times. Instead of immediately fulfilling the request, he instead hesitated and asked "don't you guys already know what this looks like?" Unbeknownst to the end-user, the context had invisibly shifted from training on one modality to training on another, resulting in his hesitation and confusion. The user and the system had two different contexts: it was not clear to him why the development team asked him to hold up the bearing. He didn't know that the context had changed, and that we were developing a new system which needed data of objects and not of tasks. This moment of confusion, friction, and hesitation led to lost time, and the need for additional needed context. Ultimately we had to explain that yes, we knew what a bearing looked like, but we were training a model using a new modality, and that we needed him to show the bearing to the camera again. Only then, once both the user and the system were established in a shared context, we were able to move forward.

\subsection{Contextual Controls.} Interactions are not bound to single contexts, however. Even within a single interaction contexts can shift from one moment to the next---with accompanying changes in interactional norms---and participants in an interaction have significant ways to control context available to them. The board meeting discussed above can shift from presentation mode to question-and-answer mode, and a presenter has control over context thanks to widely available scripts ~\cite{kellermann1992communication}. The common presentation closing, "Any questions?" accomplishes this succinctly. Similarly, human-AI interactions benefit from first making context explicit to participants, and then by having contextual controls available to participants to explicitly declare an intention to change the context. 

In our prior example of contextual integrity, the end user's hesitation and confusion made it clear contextual control needed to be provided. Developers had failed to make the contextual shift in development explicit. Following this experience, such requests were appended with explanations in the UI about what, exactly, had changed in the back-end which necessitated new behaviors on the front-end. For example, users were given messages stating "we are developing a new kind of model that needs to learn about what objects you are using. We're going to ask you to show your tools and other objects to the camera, so this new model can learn." This shifted the context from learning in one modality to learning in another, leading to a reduction in reported confusion and enabled more streamlined collaboration between the end-user and the multimodal AI system during visual bootstrapping session.

\subsection{Trust, Mistrust, \& Vulnerability.} Engaging in indeterminate interactions without certainty about the outcome requires a great deal of trust. Obeying the norms of an interaction, and particularly when surrendering oneself to interactions in which the other party can change the context seemingly at whim, all require trust. We trust that an interaction will progress in more-or-less suitable ways, that others will also follow norms, and that we will be provided with enough clues to keep up with changing contexts. Trust is a critical feature of human interactions, but is a challenging construct for human-AI interactions as it is much more difficult to establish the mutuality of interactions between humans and AI systems~\cite{burgoon2000interactivity}. Trust, in human interactions, is always accompanied by mistrust and vulnerability. In trusting another, we make ourselves vulnerable to the other~\cite{mackenzie2020vulnerability}. For designers of AI systems, recognizing the vulnerability of humans users, and also foregrounding of how vulnerable the AI system is to human interactions (who can deceive, ignore, or circumvent it) are crucial to human-AI interaction. The obverse of trust, as a way of organizing an interaction, is mistrust. A healthy skepticism can be leveraged toward a successful interaction, without requiring trust to be explicitly established~\cite{carey2017mistrust}, by providing a means to \emph{act} on mistrust. 

Designers of systems can anticipate how humans might mistrust their AI interlocutors, and provide means for humans to interrogate the AI system. For example, human mistrust grounded on the possibility of inappropriate data collection can be addressed by allowing humans to view telemetry or interaction logs, have control over or edit stored data, or otherwise generate actionable information about what data is being collected about them by the AI system. Conversely, there is room for mistrust in how AI systems address humans. By doubting the motivations of the human user and asking for clarification, a great deal of misunderstanding can be forestalled. Common misunderstandings beg clarification based on the situated context of the user~\cite{smith_socially_2004, shin_ask_2024, makitalo_talk_2002}; a query about exciting uses for fertilizer, for example, might trigger a question about why such a question is being asked (because the human user is looking for a gardening project, perhaps) before the AI system defaults to a stock safety response to any question that could reveal instructions for making harmful explosive devices. 

In another example drawn from our work with end-users of a multimodal AI assistant in the manufacturing domain, we explored their perceptions of an explainability dashboard. During that study, we explained to the end-user what an “Importance Score" meant for the automatic speech recognition model, saying “this system understands best what you're doing by aligning what you are saying out loud with what you are doing on the video. So when you use terms that mean "right now" that indicates alignment between what you're saying and what you're doing...  those kinds of time-based terms are very important to how the system understands. That's what the importance score means."
The end-user replied, "is that something I could say on purpose? Could you give me a list of phrases that would cue the system to pay attention so it can learn? If I could say "Hey watch this right now" and that would improve the AI's understanding, I would." We went beyond simple transparency - i.e. just showing the score to the end user - and embraced vulnerability, by being explicit about what the system needs help with and how end-user behaviors were understood by the system. We centered opportunities for end-users to take action. To our surprise, the end-user himself identified a new pathway for human-AI collaboration, demonstrating how constructive trust-building can be.

\subsection{Translation.}
Human interaction is rife with translational acts. We routinely rephrase things to help others interact with us, as there is not single way to represent a goal, task, desire, or thought. We often require multiple different, overlapping statements to capture our own thoughts or meaning in a way that communicates that meaning effectively. Human-AI interactions occur within systems that must translate between human and computational frames of reference, often through \emph{intermediate representations}. These representations are crucial for building robust computational models (whether they be task models, word or image embeddings, language models, or knowledge graphs~\cite{goodfellow_challenges_2013, bengio_representation_2013}) but also play a role in augmenting humans' mental models of concepts as humans and AI co-create together. 

Because such intermediate representations exist within interactions, they accommodate indeterminacy and participate in shifting interactions between contexts. Intermediate representations are translational because they provide a bridge between otherwise separate ways of describing the state of the world or specifying a goal. We deal with translational representations all the time, through imitation, specification, correction, variation, and abstraction. These are techniques at the heart of prompt engineering (see, e.g. \cite{white_prompt_2023}) in generative AI use cases, and are crucial for establishing and maintaining a context's integrity, as well as for moving toward goals and shared understandings, regardless of the degree to which they have been made explicit. 

In our work developing the manufacturing assistant, the issue of translation comes up constantly. This is particularly urgent during our participatory sessions, with a team of engineers articulating the needs for training computer vision (CV) models on one side, and the end-users attempting to train the system on the other. For example, a common challenge we face is objects obscuring the view of the primary camera collecting data to train CV models. We need to ask the end-users to move these objects out of the way. However, it can be confusing for them to understand where to move those objects, and why we're asking them to move things in their workspace. Without explicitly translating what the camera needs to see, it's common for them to move objects in the wrong direction, sometimes further obscuring the camera. We decided to translate the needs of the CV models into plainer language, and began telling the end-users that the CV models needed to see everything in a particular space in order to "understand" what was being observed. We further translated the camera's field of view into physical space by placing tape on the workbench, drawing a rectangle representing what the CV models needed to "see." This translation work enabled smoother human-AI collaboration, with far greater rates for collecting high-quality data.

\section{Conclusion}
In this paper we have drawn on key elements of human interaction that  have been robustly studied in social science to inform the study of human-computer interaction. These elements are particularly relevant for the research goals of human-AI collaboration, as demonstrated by their impact to an ongoing project developing human-AI support for technicians in a manufacturing setting. Because the elements of human-human collaboration presented here are not exhaustive, future work will develop these and incorporate additional human interaction elements, as well as empirically validate the findings that these interaction elements improve the quality of human-AI collaboration.





\bibliographystyle{acm}
\bibliography{software}

\begin{thebibliography}{10}

\bibitem{austin_how_1962}
{\sc Austin, J.~L.}
\newblock {\em How to {Do} {Things} {With} {Words}: {The} {William} {James} {Lectures} delivered at {Harvard} {University} in 1955}.
\newblock Clarendon Press, Oxford, UK, 1962.

\bibitem{barad2007meeting}
{\sc Barad, K.}
\newblock Meeting the universe halfway: Quantum physics and the entanglement of matter and meaning.
\newblock {\em Duke University Press\/} (2007).

\bibitem{liu_natural_2016}
{\sc Baraldi, S., Del~Bimbo, A., Landucci, L., and Torpei, N.}
\newblock Natural {Interaction}.
\newblock In {\em Encyclopedia of {Database} {Systems}}, L.~Liu and M.~T. Özsu, Eds. Springer New York, New York, NY, 2016, pp.~1--6.

\bibitem{bazzanella2011indeterminacy}
{\sc Bazzanella, C.}
\newblock Indeterminacy in dialogue.
\newblock {\em Language and Dialogue 1}, 1 (2011), 21--43.

\bibitem{bengio_representation_2013}
{\sc Bengio, Y., Courville, A., and Vincent, P.}
\newblock Representation {Learning}: {A} {Review} and {New} {Perspectives}.
\newblock {\em IEEE Transactions on Pattern Analysis and Machine Intelligence 35}, 8 (Aug. 2013), 1798--1828.

\bibitem{bernsen2001natural}
{\sc Bernsen, N.~O.}
\newblock Natural human-human-system interaction.
\newblock In {\em Frontiers of Human-Centered Computing, Online Communities and Virtual Environments}. Springer, 2001, pp.~347--363.

\bibitem{bernstein2023architecting}
{\sc Bernstein, M.~S., Park, J.~S., Morris, M.~R., Amershi, S., Chilton, L.~B., and Gordon, M.~L.}
\newblock Architecting novel interactions with generative ai models.
\newblock In {\em Adjunct Proceedings of the 36th Annual ACM Symposium on User Interface Software and Technology\/} (2023), pp.~1--3.

\bibitem{burgoon2000interactivity}
{\sc Burgoon, J.~K., Bonito, J.~A., Bengtsson, B., Cederberg, C., Lundeberg, M., and Allspach, L.}
\newblock Interactivity in human--computer interaction: A study of credibility, understanding, and influence.
\newblock {\em Computers in human behavior 16}, 6 (2000), 553--574.

\bibitem{carey2017mistrust}
{\sc Carey, M.}
\newblock {\em Mistrust: An ethnographic theory}.
\newblock Hau Books, 2017.

\bibitem{chalmers2003seamful}
{\sc Chalmers, M., and MacColl, I.}
\newblock Seamful and seamless design in ubiquitous computing.
\newblock In {\em Workshop at the crossroads: The interaction of HCI and systems issues in UbiComp\/} (2003), vol.~8.

\bibitem{fui-hoon_nah_generative_2023}
{\sc Fui-Hoon~Nah, F., Zheng, R., Cai, J., Siau, K., and Chen, L.}
\newblock Generative {AI} and {ChatGPT}: {Applications}, challenges, and {AI}-human collaboration.
\newblock {\em Journal of Information Technology Case and Application Research 25}, 3 (July 2023), 277--304.

\bibitem{gardner2004conversation}
{\sc Gardner, R.}
\newblock Conversation analysis.
\newblock {\em The handbook of applied linguistics\/} (2004), 262--284.

\bibitem{garfinkel2023studies}
{\sc Garfinkel, H.}
\newblock Studies in ethnomethodology.
\newblock In {\em Social Theory Re-Wired}. Routledge, 2023, pp.~58--66.

\bibitem{gaver_technology_1991}
{\sc Gaver, W.~W.}
\newblock Technology affordances.
\newblock In {\em Proceedings of the {SIGCHI} conference on {Human} factors in computing systems {Reaching} through technology - {CHI} '91\/} (New Orleans, Louisiana, United States, 1991), ACM Press, pp.~79--84.

\bibitem{glaser2017discovery}
{\sc Glaser, B., and Strauss, A.}
\newblock {\em Discovery of grounded theory: Strategies for qualitative research}.
\newblock Routledge, 2017.

\bibitem{goodfellow_challenges_2013}
{\sc Goodfellow, I.~J., Erhan, D., Carrier, P.~L., Courville, A., Mirza, M., Hamner, B., Cukierski, W., Tang, Y., Thaler, D., Lee, D.-H., Zhou, Y., Ramaiah, C., Feng, F., Li, R., Wang, X., Athanasakis, D., Shawe-Taylor, J., Milakov, M., Park, J., Ionescu, R., Popescu, M., Grozea, C., Bergstra, J., Xie, J., Romaszko, L., Xu, B., Chuang, Z., and Bengio, Y.}
\newblock Challenges in {Representation} {Learning}: {A} report on three machine learning contests, July 2013.
\newblock arXiv:1307.0414 [stat].

\bibitem{jackson_rethinking_2014}
{\sc Jackson, S.~J.}
\newblock Rethinking {Repair}.
\newblock In {\em Media {Technologies}: {Essays} on {Communication}, {Materiality}, and {Society}}, T.~Gillespie, P.~J. Boczkowski, and K.~A. Foot, Eds. MIT Press, Cambridge, MA, 2014.

\bibitem{kellermann1992communication}
{\sc Kellermann, K.}
\newblock Communication: Inherently strategic and primarily automatic.
\newblock {\em Communications Monographs 59}, 3 (1992), 288--300.

\bibitem{mackenzie2020vulnerability}
{\sc Mackenzie, C.}
\newblock Vulnerability, insecurity and the pathologies of trust and distrust.
\newblock {\em International Journal of Philosophical Studies 28}, 5 (2020), 624--643.

\bibitem{meck2024failing}
{\sc Meck, A.-M., Draxler, C., and Vogt, T.}
\newblock Failing with grace: Exploring the role of repair costs in conversational breakdowns with in-car voice assistants.
\newblock {\em International Journal of Human--Computer Interaction 40}, 22 (2024), 7574--7592.

\bibitem{meredith_analysing_2017}
{\sc Meredith, J.}
\newblock Analysing technological affordances of online interactions using conversation analysis.
\newblock {\em Journal of Pragmatics 115\/} (July 2017), 42--55.

\bibitem{makitalo_talk_2002}
{\sc Mäkitalo, A., and Säljö, R.}
\newblock Talk in institutional context and institutional context in talk: {Categories} as situated practices.
\newblock {\em Text - Interdisciplinary Journal for the Study of Discourse 22}, 1 (Jan. 2002).

\bibitem{nissenbaum_privacy_2010}
{\sc Nissenbaum, H.}
\newblock {\em Privacy in context: technology, policy, and the integrity of social life}.
\newblock Stanford Law Books, Stanford, CA, 2010.

\bibitem{obrenovic_generative_2024}
{\sc Obrenovic, B., Gu, X., Wang, G., Godinic, D., and Jakhongirov, I.}
\newblock Generative {AI} and human–robot interaction: implications and future agenda for business, society and ethics.
\newblock {\em AI \& SOCIETY\/} (Mar. 2024).

\bibitem{pidgeon1991use}
{\sc Pidgeon, N.~F., Turner, B.~A., and Blockley, D.~I.}
\newblock The use of grounded theory for conceptual analysis in knowledge elicitation.
\newblock {\em International journal of Man-machine studies 35}, 2 (1991), 151--173.

\bibitem{raudaskoski1990repair}
{\sc Raudaskoski, P.}
\newblock Repair work in human-computer interaction: a conversation analytic perspective.
\newblock In {\em Computers and conversation}. Elsevier, 1990, pp.~151--171.

\bibitem{rees_future_2025}
{\sc Rees, E.}
\newblock The {Future} of {Work}: {How} {AI} {Agents} are {Shaping} the {Next} {Wave} of {Innovation} and {Productivity} in the {Workplace}, Jan. 2025.

\bibitem{saussure_course_2011}
{\sc Saussure, F.~d.}
\newblock {\em Course in {General} {Linguistics}}.
\newblock Columbia University Press, New York, June 2011[1916].

\bibitem{shannon1948mathematical}
{\sc Shannon, C.~E.}
\newblock A mathematical theory of communication.
\newblock {\em The Bell system technical journal 27}, 3 (1948), 379--423.

\bibitem{shin_ask_2024}
{\sc Shin, J., Song, H., Lee, H., Jeong, S., and Park, J.~C.}
\newblock Ask {LLMs} {Directly}, "{What} shapes your bias?": {Measuring} {Social} {Bias} in {Large} {Language} {Models}, June 2024.
\newblock arXiv:2406.04064 [cs].

\bibitem{smith_socially_2004}
{\sc Smith, E.~R., and Semin, G.~R.}
\newblock Socially {Situated} {Cognition}: {Cognition} in its {Social} {Context}.
\newblock {\em Advances in Experimental Social Psychology 36\/} (2004), 57--121.

\bibitem{suchman1987plans}
{\sc Suchman, L.~A.}
\newblock {\em Plans and situated actions: The problem of human-machine communication}.
\newblock Cambridge university press, 1987.

\bibitem{white_prompt_2023}
{\sc White, J., Fu, Q., Hays, S., Sandborn, M., Olea, C., Gilbert, H., Elnashar, A., Spencer-Smith, J., and Schmidt, D.~C.}
\newblock A {Prompt} {Pattern} {Catalog} to {Enhance} {Prompt} {Engineering} with {ChatGPT}, Feb. 2023.
\newblock arXiv:2302.11382 [cs].

\bibitem{zittrain_we_2024}
{\sc Zittrain, J.~L.}
\newblock We {Need} to {Control} {AI} {Agents} {Now}.
\newblock {\em The Atlantic\/} (July 2024).

\end{thebibliography}

\end{document}